# New parametrization of the deuteron wave function and calculations of the tensor polarization

*V. I. Zhaba*

*Uzhgorod National University, Department of Theoretical Physics,
54, Voloshyna St., Uzhgorod, UA-88000, Ukraine
E-mail: viktorzh@meta.ua*

**Abstract**

The coefficients of new analytical forms for the deuteron wave function in coordinate space for NijmI, NijmII, Nijm93, Reid93 and Argonne v18 potentials have been numerically calculated. The obtained wave functions do not contain any superfluous knots. The designed parameters of the deuteron are in good agreement with the experimental and theoretical data. The new parameterization of the deuteron wave function in momentum space is obtained. The tensor polarization $t_{20}$ calculated based on the wave functions is proportionate to the earlier published results.

Keywords: deuteron, wave function, approximation, analytic form, polarization.

PACS: 13.40.Gp, 13.88.+e, 21.45.Bc, 03.65.Nk

## 1. Introduction

Deuteron is the most elementary nucleus, which consists of the two strongly interacting particles (a proton and a neutron). The simplicity of the deuteron's structure makes it a convenient laboratory for studying nucleon-nucleon forces. Currently, deuteron has been well investigated both experimentally and theoretically.

The experimentally determined values of static performances of the deuteron are in good agreement with the experimental data. Despite that, there still are some theoretical inconsistencies. For example, one (for Bonn potential) or both (for Moscow potential) components of the wave function have knots [1,2] near the origin of the coordinates. The presence of knots in the wave functions of the basic and sole state of the deuteron is the evidence of inconsistencies and inaccuracies in implementation of numerical algorithms in solving similar problems. The way the choice of numerical algorithms influences the solution is shown in Ref. [3, 4].

Such potentials of the nucleon-nucleon interaction as Bonn [1], Moscow [2], Nijmegen group potentials (NijmI, NijmII, Nijm93, Reid93 [5]), Argonne v18 [6] or Paris [7] potential have quite a complicated structure and cumbersome representation. The original potential Reid68 was parameterized on the basis of the phase analysis by Nijmegen group and was called Reid93. The parametrization was done for 50 parameters of the potential, where $\chi^2/N_{data}$=1.03 [5].

Besides, the deuteron wave function (DWF) can be presented as a table: through respective arrays of values of radial wave functions. It is sometimes quite difficult to operate with such arrays of numbers during numerical calculations. And the program code for numerical calculations is overloaded. Therefore, it is feasible to obtain simpler analytical forms of DWF representation. It is possible on the basis to calculate the form factors and tensor polarization, characterizing the deuteron structure.

## 2. Analytical form of the deuteron wave function

One of the first analytical forms of the deuteron wave function is forms [7]

$$\begin{cases} u(r) = \sum_{j=1}^{N} C_j \exp(-m_j r), \\ w(r) = \sum_{j=1}^{N} D_j \exp(-m_j r)\left[1 + \dfrac{3}{m_j r} + \dfrac{3}{(m_j r)^2}\right], \end{cases} \quad (1)$$

where $m_j = \beta + (j-1)m_0$, $\beta = \sqrt{ME_d}$, $m_0=0.9$ fm$^{-1}$. $M$ is nucleon mass, $E_d$ is binding energy of deuteron. The boundary conditions as $r \to 0$

$$u_b(r) \to r, \quad w_b(r) \to r^3$$

lead to one condition for $C_j$ and three constraints for $D_j$

$$\begin{cases} C_n = -\sum_{j=1}^{n-1} C_j, \\ D_{n-2} = \dfrac{m_{n-2}^2}{(m_n^2 - m_{n-2}^2)(m_{n-1}^2 - m_{n-2}^2)}\left[-m_{n-1}^2 m_n^2 \sum_{j=1}^{n-3} \dfrac{D_j}{m_j^2} + (m_{n-1}^2 + m_n^2)\sum_{j=1}^{n-3} D_j - \sum_{j=1}^{n-3} D_j m_j^2\right], \end{cases}$$

or

$$\sum_{j=1}^{N_b} C_j = 0, \quad \sum_{j=1}^{N_b} D_j = \sum_{j=1}^{N_b} D_j m_j^2 = \sum_{j=1}^{N_b} \dfrac{D_j}{m_j^2} = 0.$$

The asymptotics (1) behavior of the deuteron wave functions for large values of $r \to \infty$ are

$$u(r) \sim A_S \exp(-\beta r),$$
$$w(r) \sim A_D \exp(-\beta r)\left[1 + \dfrac{3}{\beta r} + \dfrac{3}{(\beta r)^2}\right],$$

where $A_S$ and $A_D$ are the asymptotic S- and D- state normalizations.

The Henkel transforms of wave functions $g_l(p)$ in the coordinate representation in momentum space are given by

$$g_l(p) = \sqrt{\dfrac{2}{\pi}} i^{-l} \int_0^\infty j_l(pr)\chi_l(r) dr. \quad (2)$$

The inverse Hankel transforms is

$$\chi_l(r) = \sqrt{\dfrac{2}{\pi}} i^l \int_0^\infty j_l(pr) g_l(p) dp.$$

The expressions of the momentum space for wave functions are [7]

$$\begin{cases} \dfrac{u(p)}{p} = \sqrt{\dfrac{2}{\pi}} \sum_{j=1}^{N} \dfrac{C_j}{p^2 + m_j^2} \\ \dfrac{w(p)}{p} = \sqrt{\dfrac{2}{\pi}} \sum_{j=1}^{N} \dfrac{D_j}{p^2 + m_j^2} \end{cases} \quad (3)$$

The deuteron wave functions are approximated (1) by finite sets of Yukawa-type functions (for example, DWF for the Paris [7] and CD-Bonn [1] potentials). In Ref. [8] also are present a simple parametrization as a superposition of Yukawa-type terms of the DWF, which obtained within a dispersion approach.

Except for (1), there is one more analytical form of the deuteron wave function [9]

$$\begin{cases} u(r) = A_S(1-e^{-\tau r})e^{-\alpha r}\sum_{i=0}^{n} C_i \exp(-\alpha_i r), \\ u(r) = \eta A_S(1-e^{-\sigma r})^5 k_2(\alpha r)\sum_{i=0}^{m} D_i \exp(-\alpha_i r), \end{cases} \qquad (4)$$

where $\alpha=0.2315370$ fm$^{-1}$; $\tau=5\alpha$; $\sigma=1.09$ fm$^{-1}$; $\eta=0.025$; $k_2(\alpha r)$ - terms of the spherical Bessel function:

$$k_2(\alpha r) = \left(1 + \frac{3}{\alpha r} + \frac{3}{(\alpha r)^2}\right)e^{-\alpha r}.$$

The known numerical values of a radial DWF in coordinate space can be approximated with the help of convenient expansions [10] in the analytical form (N=11):

$$\begin{cases} u(r) = \sum_{i=1}^{N} A_i \exp(-a_i r^2), \\ w(r) = r^2 \sum_{i=1}^{N} B_i \exp(-b_i r^2). \end{cases} \qquad (5)$$

For

$$u(r) = A\exp(-ar^2),$$
$$w(r) = Br^2\exp(-br^2),$$

on the Hankel transformation (2) will get the following results for the wave function in the momentum space

$$u(p) = \sqrt{\frac{2}{\pi}}\frac{A}{\sqrt{a}}F\left(\frac{p}{2\sqrt{a}}\right), \qquad (6)$$

$$w(p) = \sqrt{\frac{2}{\pi}}\frac{B}{4b^{5/2}}\frac{(12b^2 + 4bp^2 + p^4)F\left(\frac{p}{2\sqrt{b}}\right) - \sqrt{b}p(6b + p^2)}{p^2}, \qquad (7)$$

where $F(x)$ - the Dawson integral. The formulas (6) and (7) can be generalized.

To solve the system of associated Schrödinger equations that describe the radial DWF, parameterizations were proposed back in 1955 [11]:

$$\begin{cases} u = are^{-\mu r}, \\ w = bre^{-\mu r}, \end{cases}$$

$$\begin{cases} u = ar^2 e^{-\mu r}, \\ w = br^3 e^{-\mu r}. \end{cases}$$

They can be generalized for the DWF approximation as such analytical forms:

$$\begin{cases} u(r) = r^A \sum_{i=1}^{N} A_i \exp(-a_i r^a), \\ w(r) = r^B \sum_{i=1}^{N} B_i \exp(-b_i r^b), \end{cases} \qquad (8)$$

Given $N_2$=11, $a=b=3$, search for an index of function of a degree $r^n$ has been carried out, appearing as a factor before the sums of exponential terms of the analytical form (8). Best values appeared to be $n=1.47$ and $n=1.01$ for *u(r)* and *w(r)* accordingly. Hence, the factors before the sums in (8) can be chosen as $r^{3/2}$ and $r^1$ [12]

$$\begin{cases} u(r) = r^{3/2} \sum_{i=1}^{N_2} A_i \exp(-a_i r^3), \\ w(r) = r \sum_{i=1}^{N_2} B_i \exp(-b_i r^3). \end{cases} \quad (9)$$

After substitution

$$u(r) = r^{3/2} \exp(-r^3),$$
$$w(r) = r \exp(-r^3)$$

in Hankel transformation (2) get

$$u(p) = \sqrt{\frac{2}{\pi}} \frac{A}{3a^{7/6}} p \Gamma\left[\frac{7}{6}\right]_2 F_5 \left[\left(\frac{7}{12}, \frac{13}{12}\right), \left(\frac{1}{3}, \frac{1}{2}, \frac{2}{3}, \frac{5}{6}, \frac{7}{6}\right), -\frac{p^6}{11664 a^2}\right] -$$
$$- \frac{5A}{54\sqrt{2\pi} a^{11/6}} p^3 \Gamma\left[\frac{5}{6}\right]_2 F_5 \left[\left(\frac{11}{12}, \frac{17}{12}\right), \left(\frac{2}{3}, \frac{5}{6}, \frac{7}{6}, \frac{4}{3}, \frac{3}{2}\right), -\frac{p^6}{11664 a^2}\right] + \quad (10)$$
$$+ \frac{A}{240\sqrt{2} a^{5/2}} p^5 {}_2F_5 \left[\left(\frac{5}{4}, \frac{7}{4}\right), \left(\frac{7}{6}, \frac{4}{3}, \frac{3}{2}, \frac{5}{3}, \frac{11}{6}\right), -\frac{p^6}{11664 a^2}\right],$$

$$w(p) = \frac{9B}{2\sqrt{2\pi} b^{1/3}} p^{-1} \Gamma\left[\frac{7}{3}\right]_1 F_4 \left[\left(\frac{1}{6}\right), \left(\frac{1}{3}, \frac{1}{2}, \frac{5}{6}, \frac{7}{6}\right), -\frac{p^6}{11664 b^2}\right] +$$
$$+ \frac{2B\sqrt{\pi}}{3\sqrt{b}} p^{-1} \sqrt{|p|} \left[ bei_{-\frac{1}{3}}\left(\frac{2|p|^{3/2}}{3\sqrt{3}}\right) - bei_{\frac{1}{3}}\left(\frac{2|p|^{3/2}}{3\sqrt{3}\sqrt{b}}\right) - ber_{-\frac{1}{3}}\left(\frac{2|p|^{3/2}}{3\sqrt{3b}}\right) + ber_{\frac{1}{3}}\left(\frac{2|p|^{3/2}}{3\sqrt{3b}}\right) \right]$$
$$+ \frac{2B}{9b} \sqrt{\frac{2\pi}{3}} p \left[ ber_{-\frac{2}{3}}\left(\frac{2|p|^{3/2}}{3\sqrt{3b}}\right) + ber_{\frac{2}{3}}\left(\frac{2|p|^{3/2}}{3\sqrt{3b}}\right) \right] - \sqrt{\frac{2}{\pi}} \frac{B}{3b} p \, {}_1F_4\left[(1), \left(\frac{1}{3}, \frac{2}{3}, \frac{5}{6}, \frac{7}{6}\right), -\frac{p^6}{11664 b^2}\right] \quad (11)$$
$$+ \frac{B}{\sqrt{2\pi b}} p \, {}_1F_4 \left[(1), \left(\frac{2}{3}, \frac{5}{6}, \frac{7}{6}, \frac{4}{3}\right), -\frac{p^6}{11664 b^2}\right] - \frac{B}{3\sqrt{2\pi} b} p^2 \, {}_2F_5 \left[\left(\frac{1}{2}, 1\right), \left(\frac{2}{3}, \frac{5}{6}, \frac{7}{6}, \frac{4}{3}, \frac{3}{2}\right), -\frac{p^6}{11664 b^2}\right]$$
$$+ \frac{B}{90\sqrt{2\pi} b^{5/3}} p^3 \Gamma\left[\frac{2}{3}\right]_1 F_4 \left[\left(\frac{5}{6}\right), \left(\frac{7}{6}, \frac{3}{2}, \frac{5}{3}, \frac{11}{6}\right), -\frac{p^6}{11664 b^2}\right],$$

where $bei_n(x)$, $ber_n(x)$ - the Kelvin functions; ${}_1F_4[a,b,z]$, ${}_2F_5[a,b,z]$ - the hypergeometric functions. The expressions (10) and (11) are very cumbersome and complicated. It is feasible to obtain simpler analytical forms of DWF representation.

The conventional boundary conditions for radial DWF at small $r$

$$\chi_l(r) \sim r^{l+1}$$

and at $r \to \infty$

$$\chi_l(r) \sim e^{-\alpha r}$$

lead to new analytical forms in coordinate space

$$\begin{cases} u(r) = r \sum_{i=1}^{N} A_i \exp(-a_i r), \\ w(r) = r^3 \sum_{i=1}^{N} B_i \exp(-b_i r). \end{cases} \quad (12)$$

If

$$u(r) = Ar \exp(-ar),$$
$$w(r) = r^3 B \exp(-br),$$

then

$$u(p) = \sqrt{\frac{2}{\pi}} \frac{2Aap}{(a^2 + p^2)^2}, \tag{13}$$

$$w(p) = \sqrt{\frac{2}{\pi}} \frac{48Bbp^3}{(b^2 + p^2)^4}. \tag{14}$$

The formulas (13) and (14) can be generalized for the DWF approximation as such new analytical forms in momentum space

$$\begin{cases} u(p) = \sum_{i=1}^{N} \sqrt{\frac{2}{\pi}} \frac{2A_i a_i p}{(a_i^2 + p^2)^2}, \\ w(p) = \sum_{i=1}^{N} \sqrt{\frac{2}{\pi}} \frac{48 B_i b_i p^3}{(b_i^2 + p^2)^4}. \end{cases} \tag{15}$$

The accuracy of parametrization (12) is characterized by:

$$\chi^2 = \frac{1}{n-p} \sum_{i=1}^{n} \left( y_i - f(x_i; a_1, a_2, ..., a_p) \right)^2, \tag{16}$$

where $n$ - the number of points of the array $y_i$ of the numerical values of DWF in the coordinate space; $f$ - approximating function of $u$ (or $w$) according to the formulas (12); $a_1, a_2, ..., a_p$ - parameters; $p$ - the number of parameters (coefficients in the sums of formulas (12)). Hence, $\chi^2$ is determined not only by the shape of the approximating function $f$, but also by the number of the selected parameters.

Despite cumbersome and time-consuming calculations and minimizations of $\chi^2$ (to the value smaller than $10^{-7}$), it was necessary to approximate numerical values of DWF, the arrays of numbers of which made up 839x4 values in an interval $r$=0-25 fm for potentials NijmI, NijmII, Nijm93 and Reid93 [5], and 1500x2 values in an interval $r$=0-15 fm for potential Argonne v18 [6].

Based on the known DWFs (12) one can calculate the deuteron parameters (Table 1):
deuteron radius $r_d$

$$r_d = \frac{1}{2} \left\{ \int_0^\infty r^2 \left[ u^2(r) + w^2(r) \right] dr \right\}^{1/2};$$

the quadrupole moment $Q_d$

$$Q_d = \frac{1}{20} \int_0^\infty r^2 w(r) \left[ \sqrt{8} u(r) - w(r) \right] dr;$$

the magnetic moment $\mu_d$

$$\mu_d = \mu_s - \frac{3}{2}(\mu_s - \frac{1}{2}) P_D;$$

the $D$- state probability $P_D$

$$P_D = \int_0^\infty w^2(r) dr;$$

the "$D/S$- state ratio" $\eta$

$$\eta = A_D / A_S.$$

They are in good agreement with the theoretical [5,6] and experimental [13] data.

The designed DWFs (12) do not contain superfluous knots (Fig.1). They correlate well with the data in Ref. [5,6]. The values of coefficients $A_i$, $a_i$, $B_i$, $b_i$ for formulas (12) are shown in Tables 2-6.

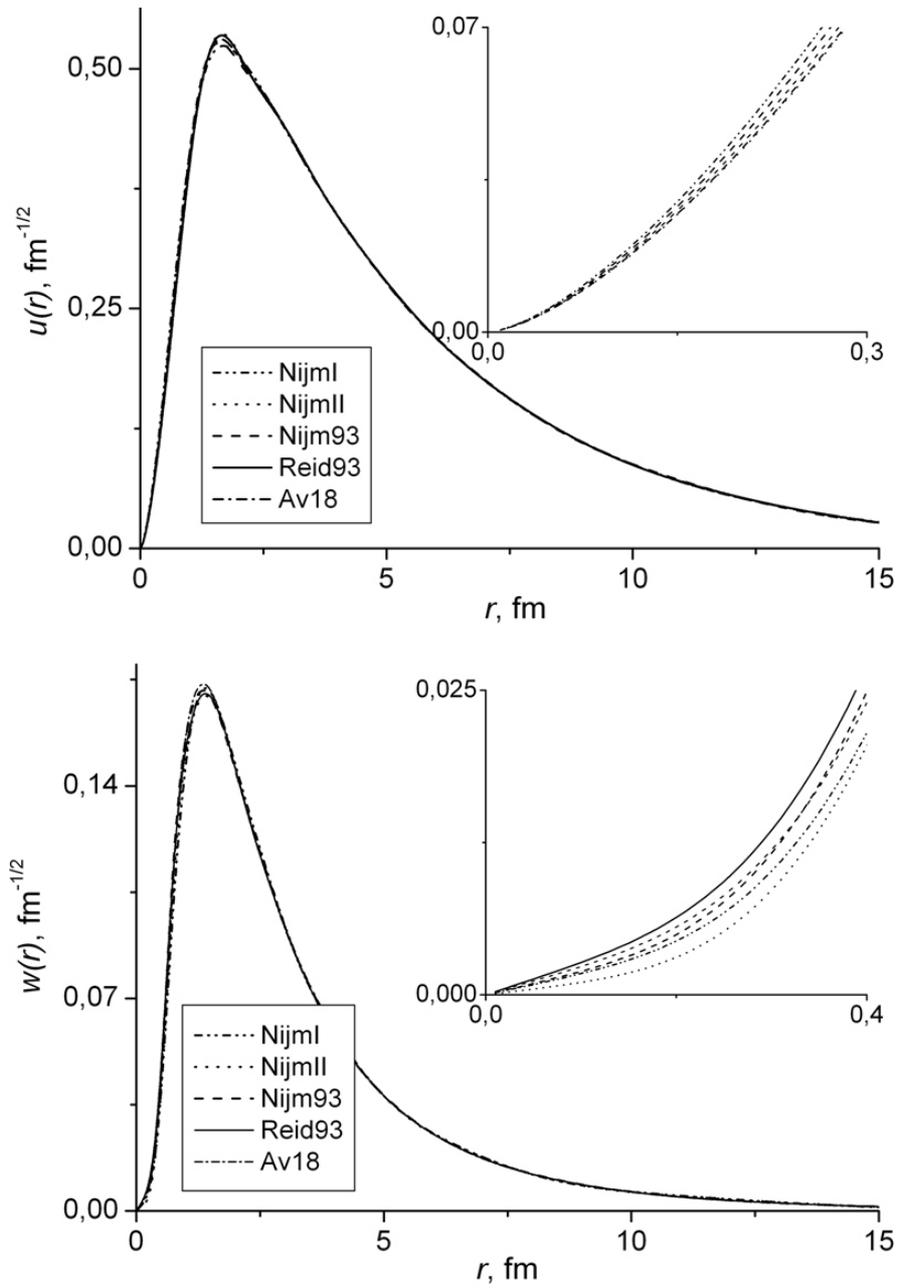

*Fig. 1.* Deuteron wave functions

*Table 1.* Deuteron properties

|  | $P_D$ (%) | $r_d$ (fm) | $Q_d$ (fm$^2$) | $\mu_d$ ($\mu_N$) | $\eta$ |
|---|---|---|---|---|---|
| Nijm I (12) | 5.651 | 1.9675 | 0.27273 | 0.847606 | 0.02545 |
| Nijm I [5] | 5.664 | 1.967 | 0.2719 | - | 0.0253 |
| Nijm II (12) | 5.633 | 1.9687 | 0.27312 | 0.847707 | 0.02542 |
| Nijm II [5] | 5.635 | 1.968 | 0.2707 | - | 0.0252 |
| Nijm 93 (12) | 5.739 | 1.9674 | 0.27003 | 0.847103 | 0.02595 |
| Nijm 93 [5] | 5.755 | 1.966 | 0.2706 | - | 0.0252 |
| Reid93 (12) | 5.700 | 1.9689 | 0.26960 | 0.847326 | 0.02483 |
| Reid93 [5] | 5.699 | 1.969 | 0.2703 | 0.8853 | 0.0251 |
| Argonne v18 (12) | 5.748 | 1.9696 | 0.27832 | 0.847049 | 0.02521 |
| Argonne v18 [6] | 5.76 | 1.967 | 0.270 | 0.847 | 0.0250 |
| Експеримент [13] | - | 1.975(3) | 0.2859(3) | 0.857438 | 0.0256(4) |

**Table 2.** Coefficients $A_i$, $a_i$, $B_i$, $b_i$ (NijmI)

| i | $A_i$ | $a_i$ | $B_i$ | $b_i$ |
|---|---|---|---|---|
| 1 | 8.34297420 | 3.11748036 | 0.07461143 | 0.82149425 |
| 2 | -8.14268678 | 2.41864844 | -11.09174655 | 0.98736190 |
| 3 | -7.97715025 | 2.44080153 | -7.36335213 | 1.13353559 |
| 4 | 2.01351965 | 1.72172981 | 1.38822344 | 1.05949390 |
| 5 | 2.02789963 | 1.72180076 | 1.38767682 | 1.05950575 |
| 6 | 1.42704820 | 1.72143867 | 1.38782848 | 1.05950037 |
| 7 | 1.49224092 | 1.72157298 | 1.38722056 | 1.05925133 |
| 8 | 0.17660683 | 0.30464720 | 1.38698258 | 1.05965487 |
| 9 | 0.96876683 | 1.72087860 | 1.37812481 | 1.05990924 |
| 10 | 1.86814770 | 1.72148573 | 1.38564435 | 1.05950877 |
| 11 | 1.80437284 | 1.72198257 | 1.39425776 | 1.05966969 |
| 12 | -0.94303743 | 1.32925623 | 1.39800077 | 1.05946552 |
| 13 | -1.01679674 | 1.32970822 | 1.36780775 | 1.02129937 |
| 14 | -0.87722585 | 1.32895799 | 1.37034932 | 1.02947430 |
| 15 | 1.15928359 | 0.80008266 | 1.41372742 | 0.90806681 |
| 16 | -0.99502549 | 1.32955650 | 1.36624049 | 0.97927761 |
| 17 | -0.94139875 | 1.32924776 | 1.48716187 | 1.93450560 |

**Table 3.** Coefficients $A_i$, $a_i$, $B_i$, $b_i$ (NijmII)

| i | $A_i$ | $a_i$ | $B_i$ | $b_i$ |
|---|---|---|---|---|
| 1 | 7.17167138 | 4.28108403 | 0.60838486 | 0.87788750 |
| 2 | -3.12798161 | 1.35213320 | -11.37934829 | 0.99065840 |
| 3 | -12.06634397 | 3.05918690 | -7.29739928 | 1.13627713 |
| 4 | 1.11021494 | 1.98616513 | 1.34605023 | 1.03020549 |
| 5 | 1.09996626 | 1.98614076 | 1.34448872 | 1.02050540 |
| 6 | 0.85731562 | 1.98502842 | 1.34487015 | 1.02351911 |
| 7 | 0.86722014 | 1.98508391 | 1.34400003 | 1.01154758 |
| 8 | 0.01091633 | 0.19622552 | 1.34875233 | 1.04086200 |
| 9 | 0.23772397 | 0.35060232 | 1.35085665 | 1.04742219 |
| 10 | 1.06879361 | 1.98602503 | 1.36766870 | 0.94005624 |
| 11 | 1.03957408 | 1.98592237 | 1.37585015 | 1.07653186 |
| 12 | 0.09800126 | 0.96224824 | 1.38093302 | 1.07645416 |
| 13 | 0.41157569 | 0.96860842 | 1.39233070 | 1.07626090 |
| 14 | 0.40038606 | 0.96949971 | 1.39192666 | 1.07626844 |
| 15 | 0.40391302 | 0.96922781 | 1.39458895 | 1.07621574 |
| 16 | 0.39674122 | 0.96965866 | 1.39340566 | 1.07624065 |
| 17 | 0.30730278 | 1.96692860 | 1.49661908 | 2.02047836 |

**Table 4.** Coefficients $A_i$, $a_i$, $B_i$, $b_i$ (Nijm93)

| i | $A_i$ | $a_i$ | $B_i$ | $b_i$ |
|---|---|---|---|---|
| 1 | 5.63304425 | 4.07423773 | -0.23822568 | 0.83937867 |
| 2 | -2.71538554 | 1.26261451 | 10.39261442 | 0.92734562 |
| 3 | -9.80059206 | 2.88656748 | -9.72030553 | 0.93288723 |
| 4 | 1.01326736 | 1.85810494 | 0.41128427 | 1.17622553 |
| 5 | 0.99911486 | 1.85806886 | 0.41141926 | 1.17638514 |
| 6 | 0.75981345 | 1.85718418 | 0.41145176 | 1.17634464 |
| 7 | 0.78614294 | 1.85732465 | 0.41007739 | 1.17585915 |
| 8 | 0.00735744 | 0.19104603 | 0.39110876 | 1.17436517 |
| 9 | 0.24562790 | 0.90075370 | -3.56756526 | 1.54815015 |
| 10 | 0.98504939 | 1.85804246 | 0.40949586 | 1.17654162 |
| 11 | 0.95074520 | 1.85795052 | 0.40393072 | 1.17435488 |
| 12 | 0.22339851 | 0.90083276 | 0.38278232 | 1.17021026 |
| 13 | 0.26701799 | 0.90091078 | -0.42991787 | 1.01243321 |
| 14 | 0.26589611 | 0.90090985 | -0.32714432 | 1.01210561 |
| 15 | 0.27037281 | 0.90091332 | -0.79488972 | 1.03789737 |
| 16 | 0.26420867 | 0.90090838 | -0.65808903 | 1.03361884 |
| 17 | 0.21458737 | 0.33547037 | 3.47374210 | 1.90421881 |

**Table 5.** Coefficients $A_i$, $a_i$, $B_i$, $b_i$ (Reid93)

| i | $A_i$ | $a_i$ | $B_i$ | $b_i$ |
|---|---|---|---|---|
| 1 | 10.82872874 | 3.57818859 | 0.85913866 | 0.84563560 |
| 2 | -3.46535514 | 1.63432788 | -11.95217749 | 0.93415140 |
| 3 | -20.37576777 | 2.77462087 | -7.00904844 | 1.01336800 |
| 4 | 1.70285401 | 1.98143484 | 1.39266529 | 0.96544809 |
| 5 | 1.70292702 | 1.98147821 | 1.39176654 | 0.96469199 |
| 6 | 1.13102396 | 1.98113296 | 1.39201276 | 0.96486913 |
| 7 | 1.16742343 | 1.98116668 | 1.39224686 | 0.96502822 |
| 8 | 0.18121123 | 0.30664177 | 1.39144392 | 0.96456376 |
| 9 | 0.81795419 | 0.77010948 | 1.38668838 | 0.92538825 |
| 10 | 1.53607432 | 1.98142813 | 1.38638328 | 0.95513199 |
| 11 | 1.46499048 | 1.98138117 | 1.37717470 | 0.97868480 |
| 12 | 1.57758672 | 1.98145128 | 1.38319409 | 0.97856519 |
| 13 | -1.27309929 | 1.63784609 | 1.39773347 | 0.97826480 |
| 14 | 1.36746946 | 1.98131766 | 1.39713320 | 0.97827828 |
| 15 | -1.16551575 | 1.63737718 | 1.40106698 | 0.97818261 |
| 16 | 1.40943904 | 1.98134467 | 1.39926402 | 0.97822895 |
| 17 | 1.61526152 | 1.98148286 | 1.41043200 | 2.28373044 |

**Table 6.** Coefficients $A_i$, $a_i$, $B_i$, $b_i$ (Argonne v18)

| i | $A_i$ | $a_i$ | $B_i$ | $b_i$ |
|---|---|---|---|---|
| 1 | -2.18820842 | 2.94118055 | 1.40149944 | 2.38647066 |
| 2 | 0.19829598 | 0.31460661 | -2.40123538 | 0.96318824 |
| 3 | 0.97998695 | 2.03171852 | 0.00288422 | 0.59397087 |
| 4 | 4.65974370 | 4.61053529 | 0.18973047 | 0.96618580 |
| 5 | -1.62522539 | 1.06465661 | 0.18975323 | 0.96618462 |
| 6 | 1.08456135 | 2.04122410 | 0.18975323 | 0.96618462 |
| 7 | 0.42156309 | 0.86645786 | 0.18486183 | 0.96641213 |
| 8 | 0.97998695 | 2.03171852 | 0.18968283 | 0.96618824 |
| 9 | 0.42156306 | 0.86647486 | 0.18975323 | 0.96618462 |
| 10 | 0.42156308 | 0.86646361 | 0.18975323 | 0.96618462 |
| 11 | -2.06552609 | 2.94167884 | 0.18968286 | 0.96618824 |
| 12 | -2.06552609 | 2.94167884 | 0.18967455 | 0.96618866 |
| 13 | 0.42156301 | 0.86650078 | 0.18973022 | 0.96618581 |
| 14 | -1.15454249 | 2.91724758 | 0.18967455 | 0.96618866 |
| 15 | -1.15454249 | 2.91724758 | 0.18969125 | 0.96618781 |
| 16 | 1.08025827 | 2.04085503 | 0.18477367 | 0.96641628 |

## 3. Form factors and tensor polarization of the deuteron

Measurement of polarization characteristics of a response of deuteron fragmentation *A(d,p)X* at the intermediate and high energies remains one of the basic tools for examination of a deuteron structure. For a quantitative understanding of the deuteron structure, *S*- and *D*- states and polarization characteristics, one should consider different models of the nucleon-nucleon potential. The deuteron charge distribution is not well known from the experiment, because it is done only through the use of polarization measurements, and the unpolarized elastic scattered differential cross sections [14-16]. However, it can be determined [14]. Differential cross section of elastic scattering of unpolarized electrons by unpolarized deuterons without measuring polarization of the repulsed electrons and deuterons [15,16]

$$\frac{d\sigma}{d\Omega} = \left(\frac{d\sigma}{d\Omega}\right)_{MOTT} S,$$

$$S = A(p) + B(p)\tan^2\left(\frac{\theta}{2}\right).$$

Here $\theta$ - the scattering angle in the laboratory system, *p* - the deuteron momentum in fm$^{-1}$, *A(p)* and *B(p)* - functions of the electric and magnetic structure [15,16]:

$$A(p) = F_C^2(p) + \frac{8}{9}\eta^2 F_Q^2(p) + \frac{2}{3}\eta F_M^2(p), \quad (17)$$

$$B(p) = \frac{4}{3}\eta(1+\eta)F_M^2(p), \quad (18)$$

where $\eta = \frac{p^2}{4M_D^2}$; $M_D$=1875.63 MeV - deuteron mass. Charge $F_C(p)$, quadrupole $F_Q(p)$ and magnetic $F_M(p)$ form factors contain information about the electromagnetic properties of the deuteron [13-17]:

$$F_C = \left[G_{Ep} + G_{En}\right]\int_0^\infty \left[u^2 + w^2\right]j_0 dr; \quad (19)$$

$$F_Q = \frac{2}{\eta}\sqrt{\frac{9}{8}}\left[G_{Ep}+G_{En}\right]\int_0^\infty\left[uw-\frac{w^2}{\sqrt{8}}\right]j_2 dr; \tag{20}$$

$$F_M = 2\left[G_{Mp}+G_{Mn}\right]\int_0^\infty\left[\left(u^2-\frac{w^2}{2}\right)j_0+\left(\frac{uw}{\sqrt{2}}+\frac{w^2}{2}\right)j_2\right]dr + \frac{3}{2}\left[G_{Ep}+G_{En}\right]\int_0^\infty w^2\left[j_0+j_2\right]dr; \tag{21}$$

where $u$, $w$ - radial DWFs (12), $j_0$, $j_2$ - the spherical Bessel functions from the argument $pr/2$; $G_{Ep}$, $G_{En}$ ($G_{Mp}$, $G_{Mn}$) - neutron and proton electric (magnetic) form factors. In the experiments with the unpolarized elastic scattering, the structure functions can be obtained by determining $B(p)$ directly from the scattering cross section back. Tensor polarization of the repulsed deuterons is determined [14-16] using the form factors (19)-(21):

$$t_{20}(p) = -\frac{1}{\sqrt{2}S}\left(\frac{8}{3}\eta F_C(p)F_Q(p)+\frac{8}{9}\eta^2 F_Q^2(p)+\frac{1}{3}\eta\left[1+2(1+\eta)tg^2\left(\frac{\theta}{2}\right)\right]F_M^2(p)\right), \tag{22}$$

$$t_{21}(p) = \frac{2}{\sqrt{3}S\cos\left(\frac{\theta}{2}\right)}\eta\sqrt{\eta+\eta^2\sin^2\left(\frac{\theta}{2}\right)}F_M(p)F_Q(p), \tag{23}$$

$$t_{22}(p) = -\frac{1}{2\sqrt{3}S}\eta F_M^2(p). \tag{24}$$

Tensor polarization (22) and (23) for NijmI, NijmII, Nijm93, Reid93 and Argonne v18 potentials (Fig. 2-4) has been carried out and the obtained results have been compared with the published experimental and theoretical data.

The impact of the accuracy of $\chi^2$ approximation (12) for DWF on calculation of tensor polarization $t_{20}(p)$ at a scattering angle of $\theta=70^0$ (Fig. 2). $t_{20}(p)$ calculated based on the two obtained approximations, has been compared. The values $\chi^2$ of these approximations for $u(r)$ and $w(r)$ make $10^{-6}$ and $10^{-6}$, $10^{-9}$ and $10^{-10}$ respectively. It appeared that the "worse" approximation significantly affects the result of calculation of $t_{20}(p)$. For example, when the momentum is 3.6 fm$^{-1}$ between approximations *1* and *2* is 2.8%.

A detailed comparison of the obtained values of $t_{20}(p)$ (the scattering angle $\theta=70^0$) for NijmI, NijmII, Nijm93, Reid93 and Argonne v18 potentials (Fig. 3) with the up-to-date experimental data of JLAB t20 [14,17] and BLAST [18,19] collaboration. There is a good agreement is for the momentas $p$=1-4 fm$^{-1}$.

The calculated value $t_{20}(p)$ is in good agreement with the results of works, where the theoretical calculations have been conducted: with data [18] for the Paris, Argonne v14 and Bonn-E potentials and with data [20] for Moscow, NijmI, NijmII, CD-Bonn and Paris potentials. It is in a good agreement with the value $t_{20}(p)$ calculated in [18] for elastic *ed*-scattering for models with the inclusion of nucleon isobaric component, within light-front dynamics and quark cluster model, for Bonn-A, Bonn-B and Bonn-C, Bonn Q, Reid-SC and Paris A-VIS potentials. There is a good agreement of the obtained values of $t_{21}(p)$ with the data for models NRIA and NRIA+MEC+RC [18]. Besides, $t_{20}(p)$ and $t_{21}(p)$ coincide well with the results according to the effective field theory [19].

The experimental data for $t_{21}(p)$ and $t_{22}(p)$ in a wide range of momentas is missing in the scientific literature. Therefore, this is of current importance to get these values both theoretically and experimentally. It is also appropriate to calculate polarization characteristics of deuteron (sensitivity tensor components to polarization of deuterons $T_{20}$, polarization transmission $K_0$ and tensor analyzing power $A_{yy}$) and compare them with theoretical calculations [4], as well as with the experimental data [21].

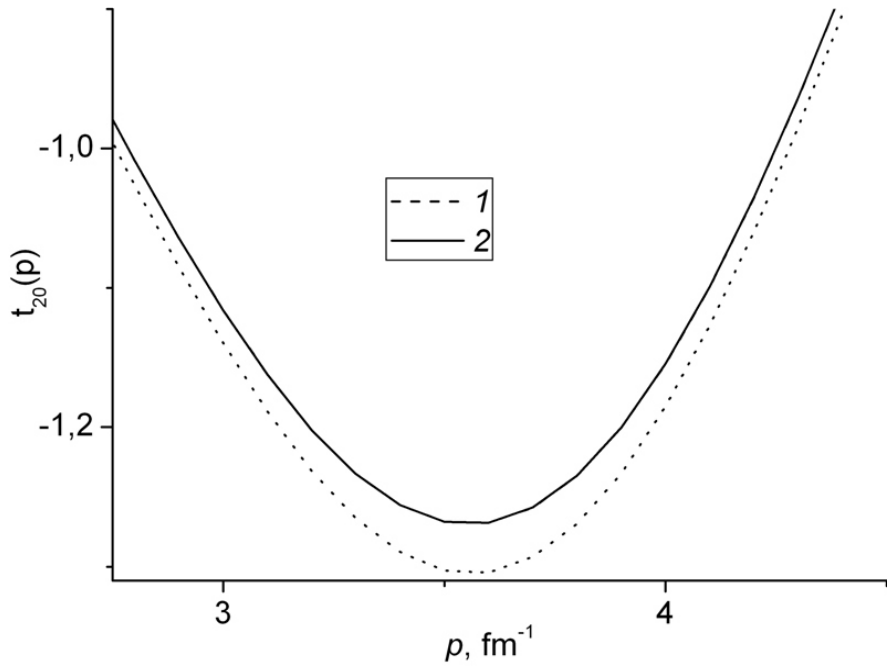

**Fig. 2.** Tensor polarization of deuteron $t_{20}$

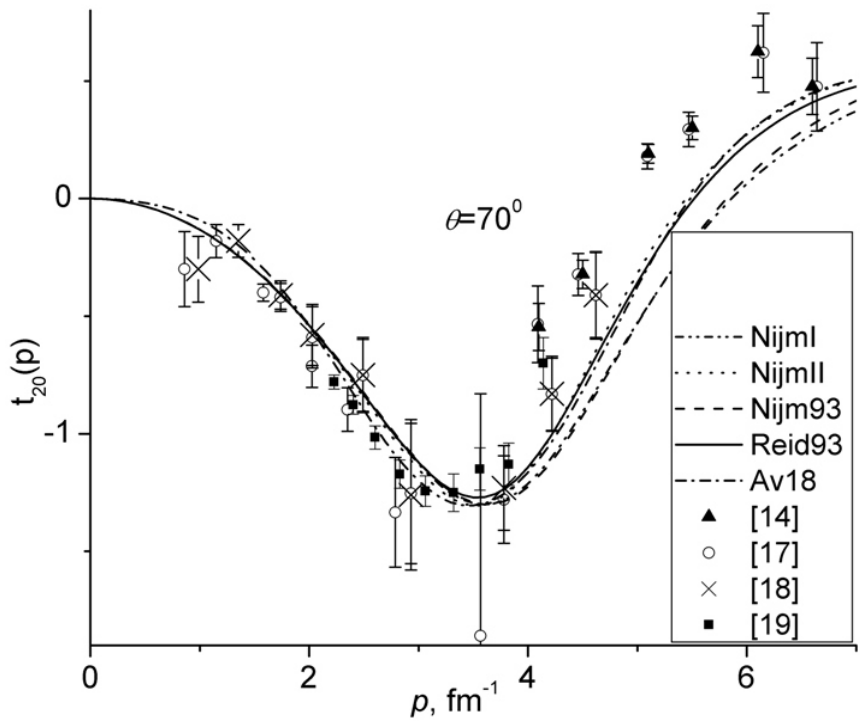

**Fig. 3.** Tensor polarization of deuteron $t_{20}$

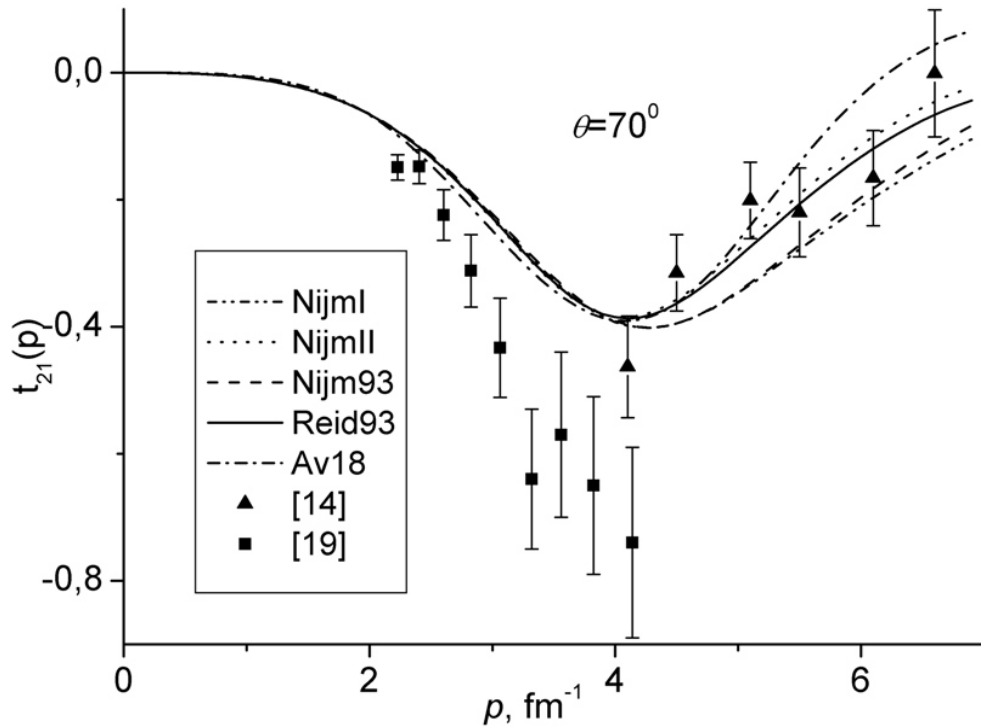

***Fig. 4.*** *Tensor polarization of deuteron $t_{21}$*

**Conclusions**

The analytic forms of the deuteron wave function in coordinate space are analyzed.

The coefficients of the approximating dependencies have been calculated in a new analytic form (12) for the numerical values of DWF in the coordinate space for realistic phenomenological potentials NijmI, NijmII, Nijm93, Reid93 and Argonne v18. With the account of the minimum values of $\chi^2$ for these forms we have built DWFs in the coordinate space, which do not contain superfluous knots. The calculated parameters of the deuteron are in good agreement with theoretical and experimental results.

The new parameterization (15) of the deuteron wave function in momentum space is obtained..

The deuteron tensor polarization has been calculated based on the received DWFs. The behavior of the value $t_{20}(p)$ depending on $\chi^2$ has been studied. Numerical calculations of the deuteron tensor polarization $t_{20}(p)$ and $t_{21}(p)$ have been carried out in the range of momentas 0-7 fm$^{-1}$. The result $t_{20}(p)$ for NijmI, NijmII, Nijm93, Reid93 and Argonne v18 potentials is in good agreement with the published results for other potential nucleon-nucleon models, as well as with the experimental data.

The obtained results of the deuteron tensor polarization $t_{ij}(p)$ give some information about the electromagnetic structure of the deuteron and differential cross section of double scattering, if there the tensor analyzing power would be known.